\documentclass[useAMS,usenatbib, usegraphicx]{mn2e}
\usepackage{lscape}
\usepackage{rotating}
\usepackage{subfigure}

\def \kms{\ifmmode{~{\rm km\,s}^{-1}}\else{~km~s$^{-1}$}\fi}
\def \vhel{\ifmmode{V_{{\rm hel}}}\else{$V_{{\rm hel}}$}\fi}
\def \vsys{\ifmmode{V_{{\rm sys}}}\else{$V_{{\rm sys}}$}\fi}
\def \degree{\ifmmode{^{\circ}}\else{$^{\circ}$}\fi}

\def \myr{\ifmmode{{\rm\ M}_\odot{\rm\ yr}^{-1}}\else{${\rm\ M}_\odot$ 
yr$^{-1}$}\fi}

\def \mdot{\ifmmode{{\rm\dot{M}}}\else{${\rm\dot{M}}$}\fi}
\def \msun{\ifmmode{{\rm\ M}_\odot}\else{${\rm\ M}_\odot$}\fi}
\def \rsun{\ifmmode{{\rm\ R}_\odot}\else{${\rm\ R}_\odot$}\fi}
\newcommand{\HA}{H$\alpha$}
\newcommand{\OIII}{[O\,{\sc iii}]\ $\lambda$5007\,\AA}

\newcommand{\NII}{[N\,{\sc ii}]\ $\lambda$6584\,\AA}

\def \apj{ApJ}
\def \mnras{MNRAS}
\def \pasp{PASP}

\def \aap{A\&A}
\def \aj{AJ}

\title[Proof of polar ejection in Abell~63]{Proof of polar ejection from the close-binary core of the planetary nebula Abell~63}

\author[Deborah L. Mitchell et al.]{Deborah L. Mitchell$^{1}$,\thanks{E-mail:
dlm@jb.man.ac.uk (DLM)} Don Pollacco$^{2}$, T. J. O'Brien$^{1}$, M. Bryce$^{1}$, J. A. L\'{o}pez$^{3}$,\newauthor 
J. Meaburn$^{1}$ and N. M. H. Vaytet$^{1}$\\
$^{1}$Jodrell Bank Observatory, School of Physics and Astronomy, University of Manchester, Macclesfield
SK11 9DL, UK
 \\
$^{2}$Department of Pure and Applied Physics, Queen's University Belfast, Belfast BT7 1NN, Northern Ireland, UK
\\
$^{3}$Instituto de Astronom{\'\i}a, Universidad Nacional Autónoma de M\'{e}xico,
 Apartado Postal 877, 22800 Ensenada, B.C., M\'{e}xico
}
\begin{document}

\date{Accepted ... Received...}

\pagerange{\pageref{firstpage}--\pageref{lastpage}} \pubyear{2006}

\maketitle

\label{firstpage}

\begin{abstract}

We present the first detailed kinematical analysis of the planetary nebula Abell~63, which is known to contain the eclipsing close-binary nucleus UU~Sge. Abell~63 provides an important test case in investigating the role of close-binary central stars on the evolution of planetary nebulae.

Longslit observations were obtained using the Manchester echelle spectrometer combined with the 2.1-m San Pedro Martir Telescope. The spectra reveal that the central bright rim of Abell~63 has a tube-like structure. A deep image shows collimated lobes extending from the nebula, which are shown to be high-velocity outflows. The kinematic ages of the nebular rim and the extended lobes are calculated to be 8400$\pm$500 years and 12900$\pm$2800 years, respectively, which suggests that the lobes were formed at an earlier stage than the nebular rim. This is consistent with expectations that disk-generated jets form immediately after the common envelope phase. 

A morphological-kinematical model of the central nebula is presented and the best-fit model is found to have the same inclination as the orbital plane of the central binary system; this is the first proof that a close-binary system directly affects the shaping of its nebula. A Hubble-type flow is well-established in the morphological-kinematical modelling of the observed line profiles and imagery.

Two possible formation models for the elongated lobes of Abell~63 are considered (1) a low-density, pressure-driven jet excavates a cavity in the remnant AGB envelope; (2) high-density bullets form the lobes in a single ballistic ejection event. 
\end{abstract}

\begin{keywords}
circumstellar matter -- stars: mass-loss -- stars: winds, outflows -- stars: 
kinematics -- ISM: planetary nebulae: individual: Abell~63.
\end{keywords}

\section{Introduction}

Understanding the shapes of planetary nebulae (PNe) has proved elusive. Whilst it is
generally accepted that a large density contrast between the
equatorial and polar regions is needed to produce the extreme
morphologies of bipolar nebulae, no consensus has arisen as to its
nature. The leading theories involve interaction from either (or both)
a magnetic field or a binary companion. While there may be some evidence
for an enhanced binary fraction amongst central stars, this is far from
certain (Sorrensen \& Pollacco 2004, De Marco et al. 2004).

Until recent years, the Generalised Interacting Stellar Winds (GISW) model has been the accepted model to explain the formation of bipolar PNe (Kahn \& West 1985, Balick et al. 1987). The GISW model assumes that the red giant wind is preferentially dense in the equatorial plane, and the fast wind is subsequently focused in the polar direction. Common envelope ejection from a close-binary system provides a possible mechanism to produce the required density contrast between the equatorial and polar regions (Morris 1987); however, the ratio of this density contrast has never been established. 

It was noted by \cite{2000ApJ...538..241S} that the GISW model could only account for PNe with prolate or mildly bipolar structures. The model failed to reproduce those PNe with very narrow waists or collimated, polar outflows, e.g. M~2-9, Mz~3 and HB~12. \cite{1990AJ.....99.1869S} attempted to reproduce the ansae commonly observed in prolate and bipolar PNe using numerical simulations of a two-wind interaction, but was unable to produce sufficient focusing of the fast wind towards the symmetry axis. \cite{1998AJ....116.1357S} argued that high-speed collimated outflows or jets were responsible for producing highly collimated PNe instead of a pre-existing equatorial density enhancement. \cite{1994ApJ...421..219S} claimed that a stellar accretion disk is required to produce collimated outflows, which infers the presence of a binary companion. 

Simulations have shown that as a close-binary system undergoes a common
envelope phase, if the companion star accretes matter with a high enough angular momentum, an accretion disk is formed (Soker 2002). Hence, as the evolving star moves to
higher temperatures, its wind will blow out through the areas of
lowest density, i.e. the polar regions (Morris 1987).

\cite{1998ApJ...497..303M} investigated the effects of a detached binary star on a mass-losing giant star. In all of their models accretion disks formed around the binary companion. \cite{1999ApJ...524..952R} investigated the types of binary system that would form accretion disks as a consequence of common envelope evolution. They found that an accretion disk will only form if the orbital separation is less than 2\rsun\ and the secondary mass is less than 0.08\msun.   

In recent years many authors have proposed magnetic fields to be crucial in
shaping PNe. In general, these models assume a fast, rotating 
core of the precursor AGB star. While this is needed to generate the magnetic field, it 
has proved difficult to model how such fast rotation of the star could arise.  One possibility is 
that the core has been spun up through the interaction with a close companion, presumably during the common envelope phase. Polarisation measurements
have been interpreted as indicating the existence of fields (Vlemmings et al.~2006) but this 
explanation is not universally accepted (Soker 2006). 

It is also possible that jets produced by an accretion disk may help shape the nebula.
In these models, the disk could be around either component and the jets could be launched
in a magnetic (Frank 2005) or non magnetic process \citep{2004AA...422.1039S}.
In either case, the presence of the disk implies the existence of a companion star. The jets will blow cavities through the AGB envelope along the polar axis of the binary. Magnetic confinement models predict a linear increase in the expansion velocity of the jet with distance from the centre of the nebula (Garc{\'{\i}}a-Segura et al. 1999). 

Some central stars of PNe are well known
close-binaries (Bond \& Livio 1990, Pollacco \& Bell 1997), three in particular are actually eclipsing systems. These
objects are especially important as their physical parameters can be
determined, at least in principle, in a model independent way; hence
their distances can be derived with far greater accuracy than is
common for PN central stars. Consequently, the properties of these
objects can be used as tests of stellar evolution theory.

The eclipsing central stars are Abell 46, Abell 63 and SuWt2. Whilst
the latter object appears peculiar in many respects, the Abell objects
are typical of old, low surface brightness objects.  A provisional
analysis of the binary orbit and light curve of V477 Lyr, the central star of Abell~46
\citep{1994MNRAS.267..452P}, shows the system to be partially eclipsing 
and the secondary component to be of extremely low mass.  UU~Sge is the central star of
Abell~63, and  was discovered by \cite{1978ApJ...223..252B} to be eclipsing and they derived
physical parameters based on the light curve alone. More recent analysis has
shown the eclipse of the primary to be total, lasting for some twelve
minutes, and has enabled physical parameters to
be derived with unprecedented accuracy (Bell et al. 1994, Pollacco \& Bell 1993). 

A deep \HA\ + \NII\ image of Abell~63 is shown in Fig.~1 (taken from Pollacco \& Bell 1997). Abell~63 has a bright nebular rim (shown at high-contrast inset) and from this, faint, well-confined extensions are visible, which lead
to ``end caps'' (labelled A and B in Fig.~1) giving the overall impression of a tube with a
waist. The aspect ratio of the tube is about 7:1 from end to end and in many respects
is similar to the extreme bipolar objects M2-9 and Mz3. The overall dimensions of the tube are 290 $\times$ 42 arcsec$^{2}$. The end caps are approximately equi-distant from the centre of the nebula.  Imaging presented by \cite{1997MNRAS.284...32P} show the end caps to be brightest in \NII. A faint shell is visible surrounding the bright nebular rim and its edges are clearly defined by a sharp drop in brightness. This is likely to be the relic of the ionised AGB wind. The nebula is also expected to be surounded by a faint red giant halo, although this has not been observed. 

Until now, there has been no observational proof that a central binary star directly affects the shaping of its nebula or is responsible for the formation of a jet-like structure. In the case of Abell 63, the
central binary, UU Sge, has been well characterised and hence, at least in
principle, allows direct testing of the common envelope hypothesis.

Surprisingly, very little observational work has been carried out on PNe containing close-binary central stars. Thus, in the present paper the kinematics of Abell~63 are explored with a view to investigating the kinematical and morphological effects of binarity.

\section[]{Observations and Data Reduction}

\begin{figure*}
\centering
\mbox{\resizebox{11.0cm}{!}{\includegraphics{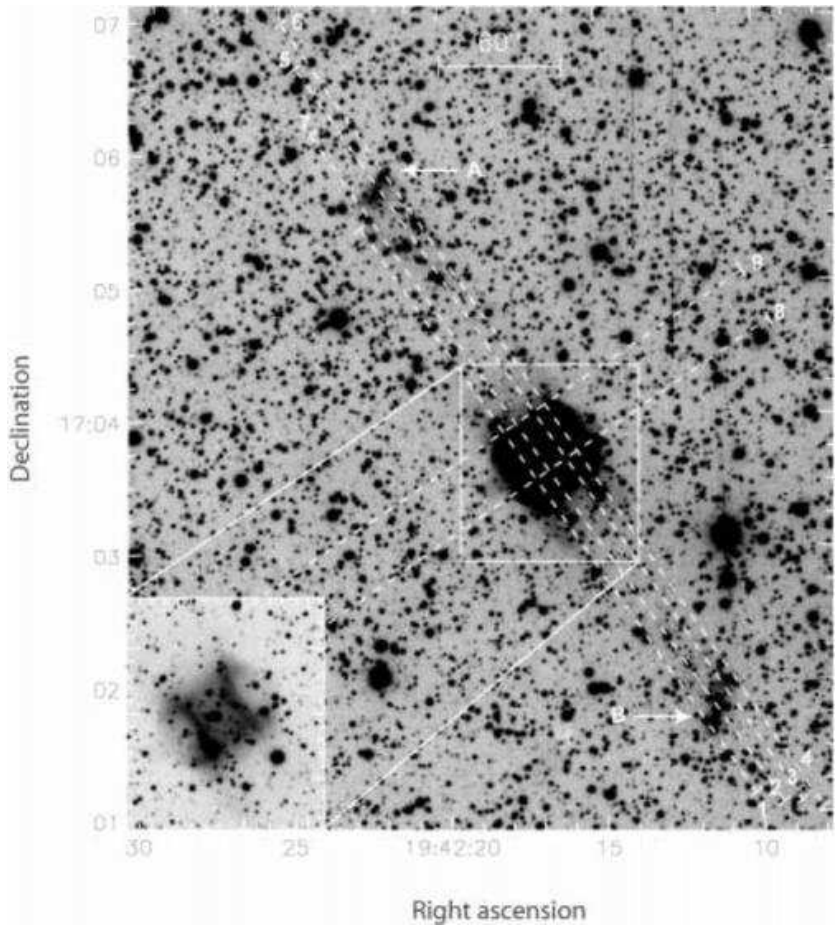}}}
\caption {A narrow-band \HA\ + \NII\ image of Abell~63 taken from Pollacco \& Bell (1997). The large image shows the full extent of the nebula at low-contrast. Slit positions are drawn on the image. Note that slit positions 2, 3 and 4 overlap with the displaced positions 5, 7 and 6, respectively. Two end-caps are visible at the tips of the collimated lobes, which are labelled A and B. The insert shows the bright central nebula at high contrast.}
\end{figure*}

Spatially resolved, longslit emission line profiles
were obtained of Abell~63 at high spectral resolution with the second Manchester echelle spectrometer combined with the 2.1-m San Pedro M\'{a}rtir telescope (MES-SPM) (Meaburn et al. 2003). MES-SPM was used in its primary spectral mode with a narrow-band 90 \AA\ filter to isolate the \HA\ and [N\,{\sc ii}]\ 6548 and 6584~\AA\ emission lines in the
87th echelle order. A deep image was also taken in the light of \OIII; however, the nebula appeared very faint and very little structure was resolved compared with the \HA\ observations. Observations took place in June 2004 using a SITe3 CCD with 1024$\times$1024 24 $\mu$m square pixels ($\equiv$ 0.31\arcsec pixel$^{-1}$). All integrations were of 1800-s duration.

Binning of 2 $\times$ 2 was adopted during the observations, giving
512 pixels in the spectral (x) direction ($\equiv$ 4.79 kms$^{-1}$
pixel$^{-1}$) and 512 pixels in the spatial (y) direction ($\equiv$
0.62\arcsec pixel$^{-1}$). This gave a projected slit length of 5.3
arcmin on the sky. The slit was 150 $\mu$m wide ($\equiv$ 2.0\arcsec\
and 10 kms$^{-1}$).

The slit positions are drawn on the deep image of Abell~63 shown in Fig.~1. Seven integrations were obtained with the slit orientated along
the major axis of the nebula at PA = 34$^{\circ}$; as the slit was not long
enough to cover the whole extent of the nebula between the end-caps, three slit
positions covered  end-cap A and four covered end-cap B, allowing sufficient overlap of the slits across the bright central nebula. Two integrations
were made with the slit orientated across the minor axis of the central nebular shell at PA = -56$^{\circ}$.

Data reduction was performed using \textsc{starlink} software. The
spectra were bias-corrected and cleaned of cosmic rays. The
spectra were then wavelength calibrated against a ThAr emission line
lamp.

\section[]{Results}

\begin{figure*}
\centering
\mbox{\resizebox{8.0cm}{!}{\includegraphics{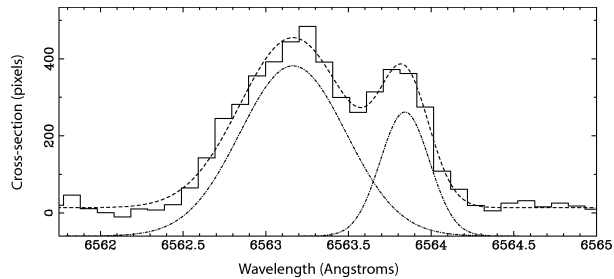}}}
\caption {An example of a double Gaussian fit to an \HA\ emission profile from the bright central nebula (slit position 3).}
\end{figure*}

\begin{figure*}
\centering
\mbox{\resizebox{17.0cm}{!}{\includegraphics{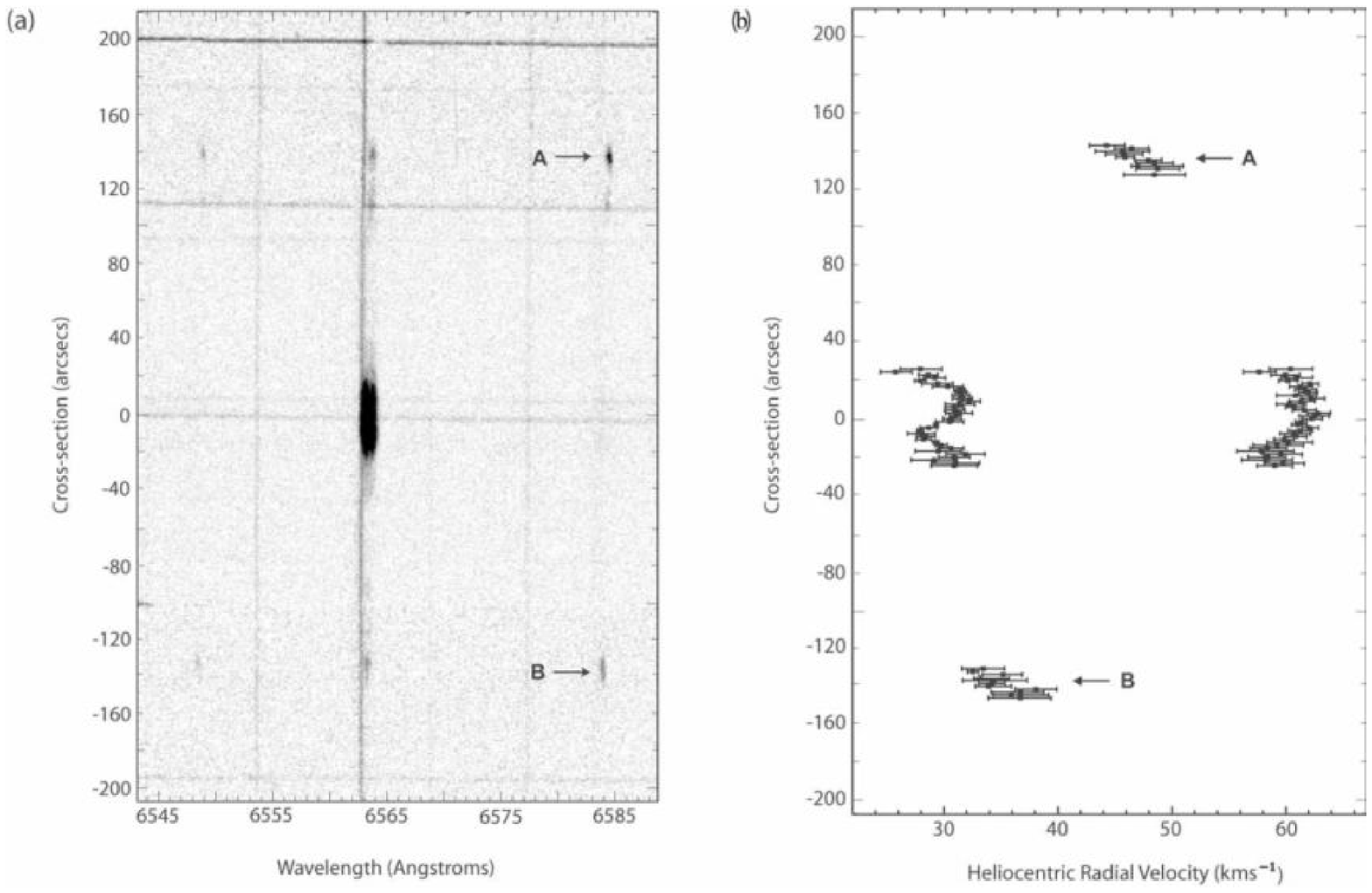}}}
\caption {(a) An \HA\ + \NII\ longslit spectrum showing emission from the bright rim of Abell~63 and both end-caps (labelled A and B - see Fig.~1). This is a composite spectrum from slit positions 2 and 7 that overlap along the central nebula. The rim is bright in \HA\ and the end-caps are visible in both \HA\ + \NII. The lobes are faintly visible in \HA\ and extend between the bright central nebula and the end-caps. Note that the stellar continuum lines are not quite horizontal in this spectrum as a result of a slight misalignment between the CCD axis and the echelle grating. (b) Centroids of best-fit Gaussians to velocity components in observed \HA\ + \NII\ profiles ($\equiv$ \vhel) from an extended slit (positions 2 and 7) aligned along the major axis of Abell~63. The positions of the end-caps are labelled A and B, respectively}
\end{figure*}

\begin{figure*}
\centering
\mbox{\resizebox{8.0cm}{!}{\includegraphics{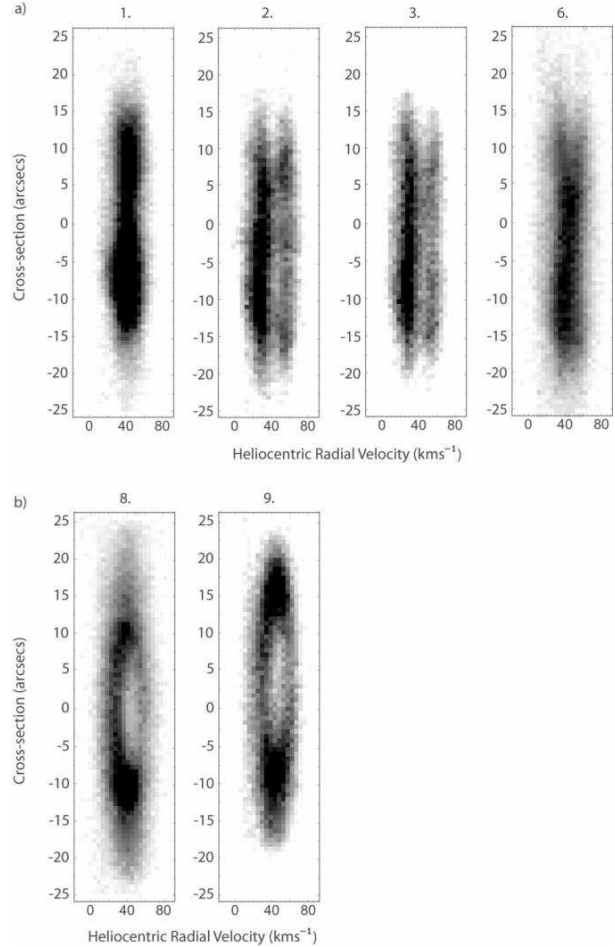}}}
\caption {Longslit \HA\ spectra from the bright rim and faint, surrounding shell of Abell~63. Longslit spectra from slit positions 1, 2, 3 and 6, which are aligned along the major axis of the nebula, are shown in (a) and longslit spectra from perpendicular slit positions 8 and 9 are shown in (b).}
\end{figure*}

\begin{figure*}
\centering
\mbox{\resizebox{10.0cm}{!}{\includegraphics{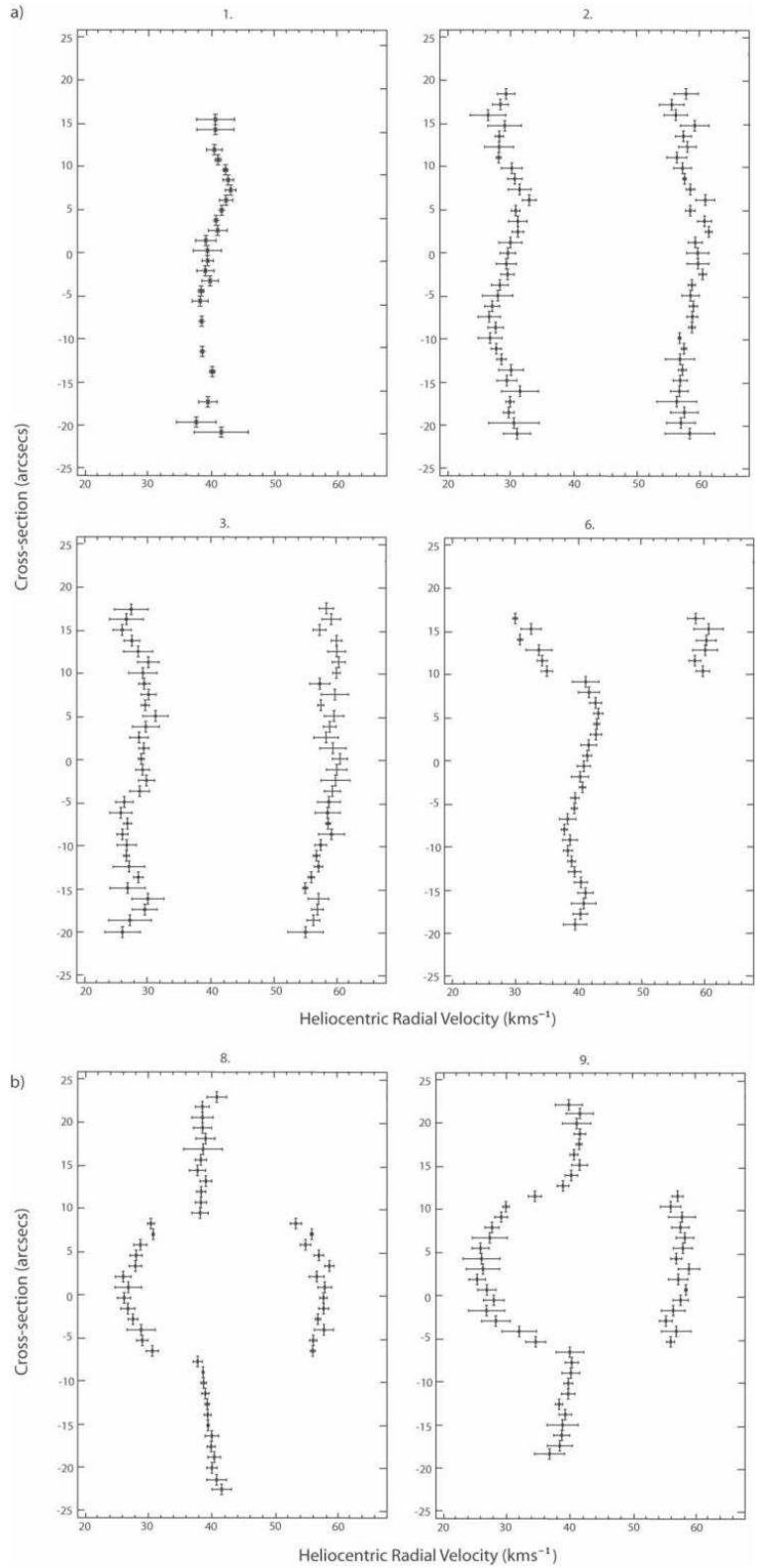}}}
\caption {Centroids of best-fit Gaussians to velocity components in observed \HA\ profiles ($\equiv$ \vhel) from the bright rim and faint surrounding shell of Abell~63. Plots of \vhel\ are shown from slit positions 1, 2, 3 and 6 in (a) and plots of \vhel\ from slit positions 8 and 9 are shown in (b).}
\end{figure*}

\begin{figure*}
\centering
\mbox{\resizebox{8.0cm}{!}{\includegraphics{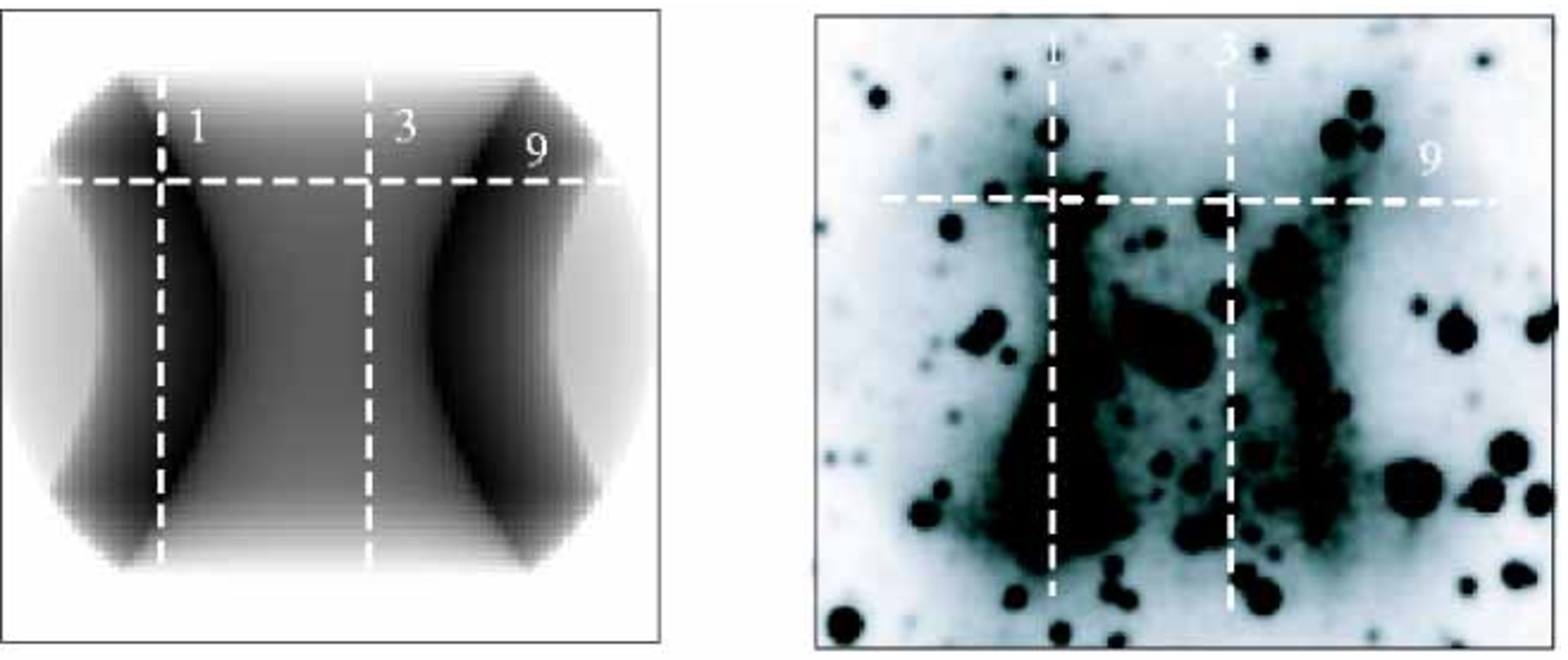}}}
\caption {A comparison of the morphological-kinematical model of Abell~63 (left) that best explains the observed structure with a direct image of the central nebula (right). Slit positions 1, 3 and 9 are drawn on both the real image and model.}
\end{figure*}

\subsection[]{Longslit spectra}

The manual fitting program \textsc{longslit} was used to fit
Gaussian profiles to individual velocity components within each \HA\ and \NII\ line profile. The profiles were binned together into blocks of two in order to improve the signal-to-noise ratio in each profile.

It was only possible to perform reliable Gaussian fitting to the bright \HA\ emission from the inner nebula (including the faint shell for slit positions 8 and 9) and the \NII\ emission from the end-caps; the emission from the extended lobes was too faint to perform Gaussian fitting in this region. Fig.~2 shows an example of a Gaussian fit to an \HA\ emission profile from the bright rim of Abell~63. \textsc{longslit} was used to produce
plots of heliocentric radial velocity (\vhel) of the fitted Gaussians centroids as a function of the slit length. The line profiles are
calibrated to $\pm$1 kms$^{-1}$ in absolute heliocentric velocity. 

An \HA\ + \NII\ longslit spectrum showing the full spatial extent of the major axis of Abell~63 is shown in Fig.~3a (adjoining longslit spectra from vertically overlapping slit positions 2 and 7). The corresponding plot showing the variation of \vhel\ along the slit is shown in Fig.~3b. In Figs.~3a and b, zero arcsecs corresponds to the position of the central star of the nebula. 

The nebular rim is bright in \HA\ and the surrounding shell also exhibits faint \HA\ emission, but neither are visible in \NII\ (Fig.~3a). The end-caps are bright in \NII\ and are also faintly visible in \HA. Fig.~3b shows that the heliocentric radial velocities of the end caps are ${V_{{\rm sysA}}}$ = 47 kms$^{-1}$ and ${V_{{\rm sysB}}}$ = 36 kms$^{-1}$, each with a range of 2 kms$^{-1}$.  Faint material from the lobes between the inner nebula and the end-caps is also visible in \HA. 

Fig.~4a shows subsets of the longslit spectra from slit positions 1, 2, 3 and 6, which are positioned along the major axis of Abell~63 (Fig.~1). Only \HA\ emission from the bright nebular rim and faint, surrounding shell is shown for each slit position in the range $\pm$25 arcsecs in order to reveal small-scale variations in the kinematics. Slit positions 4, 5 and 7 are not shown because they overlap with positions 6, 3 and 2, respectively along the central nebular shell (Fig.~1), so the longslit spectra in this overlapping region are identical. Fig.~4b shows subsets of the longslit spectra from slit positions 8 and 9, which are orientated across the minor axis of the central nebular shell (Fig.~1). The corresponding plots of \vhel\ for slit positions along the major and minor axes of the central nebula are shown in Figs.~5a and 5b, respectively. 

Slit 1 is positioned along the eastern edge of the nebula (Fig.~1). The corresponding longslit spectrum (Fig.~4a) shows a single \HA\ component from the nebular rim between $\pm$15 arcsecs. There is little variation in \vhel\ along the major axis of the central nebula at this slit position and the average \vhel\ is 39 kms$^{-1}$ (Fig.~5a).
   
Slit position 2 is offset by 5 arcsec to the east of the central star (Fig.~1). The bright \HA\ component shows splitting along the entire major axis of the nebular rim (Fig.~4a) of $\sim$ 28 kms$^{-1}$ (Fig.~5a). Both blue- and red-shifted components show little variation in \vhel\ with distance along the slit.

Slit position 3 is aligned 5 arcsec to the west of the central star (Fig.~1). The longslit spectrum (Fig.~4a) and \vhel\ plot (Fig.~5a) are almost identical to those observed for slit position 2, whereby an \HA\ line-splitting of $\sim$28 kms$^{-1}$ is observed along the bright central nebula. Again, there is very little variation in velocity along the central nebula and the same ``Z-shape'' trend in \vhel\ is apparent. These slight shifts in velocity are probably real kinematic features as they are observed in the longslit spectra from each side of the central star.
  
Slit 6 is aligned along the western edge of the nebular rim (Fig.~1). The longslit spectrum exhibits a single \HA\ component (Fig.~4a) that deviates very little from 40 kms$^{-1}$ apart from a line-splitting of 28 kms$^{-1}$ at the top edge of the nebular rim (between 8 - 17 arcsecs, Fig.~5a).   

Slits 8 and 9 are aligned across the minor axis of Abell~63, perpendicular to all other slit positions. Slit 8 is positioned 10 arcsec south of the central star and slit 9 is 20 arcsec to the north. Both slit positions intersect with the faint nebular shell surrounding the bright rim. The longslit spectra (Fig.~4b) are both velocity ellipses with a maximum line-splitting of 34~kms$^{-1}$ (Fig.~5b).  The blue-shifted component of the ellipse from slit 8 is brighter than the red-shifted component; however, both red- and blue-shifted sides of the ellipse have equal brightness in the longslit spectrum from slit 9. 

Line splitting is not observed in the emission from the faint shell surrounding the bright rim; however, some line broadening could be expected if it is a thick expanding shell. 

\subsection[]{The systemic velocity of Abell~63}

As this work presents the first kinematical study of Abell~63, there has been no previous determination of its systemic heliocentric radial velocity, \vsys. \vsys\ can be simply determined from the variation in \vhel\ across the minor axis of Abell~63 (Fig.~5b). If the nebula was at an inclination of 90$^{\circ}$ to the line-of-sight, then the expansion velocity at the edges of the tube would be perpendicular to the line-of-sight; thus the only observable component of \vhel\ would be \vsys\ of the nebula; however, assuming that this is not the case and Abell~63 has the same inclination as the orbital plane of the binary system, 87.5$^{\circ}$ (see \S4), then we will observe a small component of the expansion velocity of the tube in addition to \vsys. In this case the velocity ellipse will appear slightly tilted, whereby one side of the ellipse is blue-shifted and the other is red-shifted relative to \vsys, as seen in Fig.~5b. Assuming that \vhel\ exhibits equal red- and blue-shifts from \vsys\ at each side of the velocity ellipse, then we deduce \vsys\ for Abell~63 is 41$\pm$2 kms$^{-1}$.

The kinematics of the end-caps provide an alternative, independent method for determining \vsys\ of the nebula. It is assumed that the end-caps are expanding at equal velocities radially away from the central nebula. The observed difference in \vhel\ between the end-caps (Fig.~3b) is due to the inclination of the system, i.e. if the nebula was at 90$^{\circ}$ to the line-of-sight, the only observable component of \vhel\ from the end-caps would be \vsys. At an inclination less than 90$^{\circ}$, we observe one of the end-caps to be red-shifted relative to \vsys\ and the other to be blue-shifted. It follows that \vsys\ lies halfway between the measured values of \vhel\ for end-caps A and B, which gives \vsys\ = 41.5$\pm$2 kms$^{-1}$. This is consistent with the estimate of \vsys\ obtained from the kinematics of the central nebula.

\subsection{Kinematic ages of the nebular shell and end-caps}

The observed transverse expansion velocity of the nebular rim is approximately constant at 17$\pm$1~kms$^{-1}$. The expansion velocity is found to be consistent for longslit spectra from both the major and minor axes of Abell~63. Assuming axial symmetry and using the accurately known distance to Abell~63 of 2.4 kpc (Pollacco \& Bell 1993) then the minor axis of the tube-like rim is measured to be 0.15~pc. Assuming that the expansion of the nebular rim has been constant over the lifetime of the nebula, its kinematic age is found to be 8400$\pm$500 years.

Although the \HA\ line provides a bulk velocity across the rim, we point out that the turbulent width of the \HA\ line is in the order 12 kms$^{-1}$ and thus there is no evidence of significant velocity differentials across the nebular rim.

The expansion velocity of the lobes along the major axis is measured at the bright end-caps A and B. As stated in \S3.1, end-caps A and B are expanding away from the central nebula with an average \vhel\ of $\sim$47~kms$^{-1}$ and $\sim$36 kms$^{-1}$, respectively (Fig.~3b). Subtracting \vsys\ of the nebula gives a radial velocity difference of 5.5$\pm$1~kms$^{-1}$. Assuming that the nebula has the same inclination as the orbital plane of the central binary system, 87.5$^{\circ}$, then this gives a true expansion velocity of 126$\pm23$ kms$^{-1}$. The angular separation of the endcaps is 284.7 arcsec (Fig.~1), which represents a physical distance of 3.31 pc. Thus the propagation distance for an individual end-cap is 1.66 pc. The kinematical age of the lobes is then 12800$\pm$2800 years. The large error in the kinematical age arises because the lobes are expanding in the plane of the sky. Our analysis shows that the lobes are older than the nebular rim by at least 1100~years. 

Of course, the derived kinematical age of the rim should be used with caution in this way as the rim is expected to steadily accelerate over time under the increasing pressure of the thermalised stellar wind; this effect was demonstrated in numerical simulations by \cite{2005A&A...441..573S}. In the case of a star with a mass in the range 0.595 - 0.605\msun\ and a true age of 10000 years, their simulations predict a discrepancy of $\sim$ 1500 years between the true age and kinematic age of the rim; however, the simulation did not take into account the effects of axisymmetry or clumpiness in the nebular structure and did not account for binarity of the central star.
 
\section{A morphological-kinematical model for Abell~63}

The present kinematical observations support our hypothesis that Abell~63 has a tube-like structure; the \HA\ line-splitting observed along the major axis of the nebular rim (Fig.~5a) is consistent with observing a hollow tube along its major axis, whereby the red- and blue-shifted line components emanate from the back and front sides of the expanding tube, respectively. As the split velocity components do not converge, this suggests the tube is open-ended. 

The longslit spectra from slits 8 and 9 (Fig.~5b), which are parallel to the minor axis of the nebula, show velocity ellipses. This is consistent with viewing a radially-expanding hollow tube with circular section in cross-section. 

Slits 1, 4 and 6 are positioned along the edges of the nebular rim. In each case, line-splitting is only observed towards the top of the rim, and most
 of the emission can be fitted with a single Gaussian profile. This reflects
 the thickness of the tube - if the slit does not intersect with the hollow
 centre of the tube then no line splitting is expected. 

We have
 constructed the most simple three-dimensional morphological-kinematical model that is consistent with the observed
 large-scale velocity features of Abell~63 (Fig.~5). This model does not contain any dynamical explanation for the structure or dynamics of the nebula, it simply explains the three-dimensional structure and radial velocities observed. The model was created using the code described in \cite{1999MNRAS.307..677G}, which produces a grid consisting of two spatial dimensions in the plane of the sky and one dimension corresponding to velocity along a line-of-slight perpendicular to this plane. The model assumes axisymmetry and allows for arbitrary inclination to the line-of-sight. A function is input into the 2-dimensional grid to represent the shape of the nebula and then the line-of-sight velocity is calculated for each element of the grid. The datacube is then collapsed along the velocity axis to produce a 2-dimensional image of the model. 

Only the bright rim and surrounding, faint shell of Abell~63 were modelled (the region of the nebula shown in the right-hand panel of Fig.~6) as the lobes are too faint to obtain kinematical information. The observed shape of the nebular rim was created using a hyperbolic
 function multiplied by a
 stretching factor to determine how quickly the maximum radius was reached (this determines how narrow the waist is). The measured height-to-width ratio of the central region of Abell~63 has been preserved in the model. In order to test the hypothesis that Abell~63 has the same inclination as the orbital plane of its binary system, we used model inclinations of 87.5$^{\circ}$ (the inclination of the central binary system, UU Sge), 80.0$^{\circ}$ and 70.0$^{\circ}$.

\begin{figure}
\centering
\mbox{\resizebox{7.0cm}{!}{\includegraphics{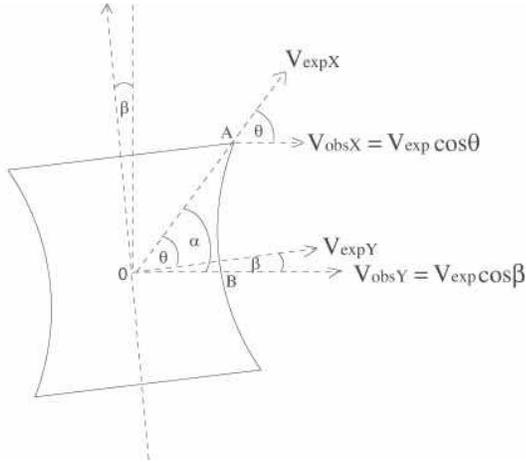}}}
\caption {A schematic of the morphological-kinematical model at an inclination $\beta$ to the line-of-sight. The expansion velocity of the model is directly proportional to the radius from point 0, i.e. the model obeys a Hubble-type flow.}
\end{figure}

\begin{figure}
\centering
\mbox{\resizebox{6.5cm}{!}{\includegraphics{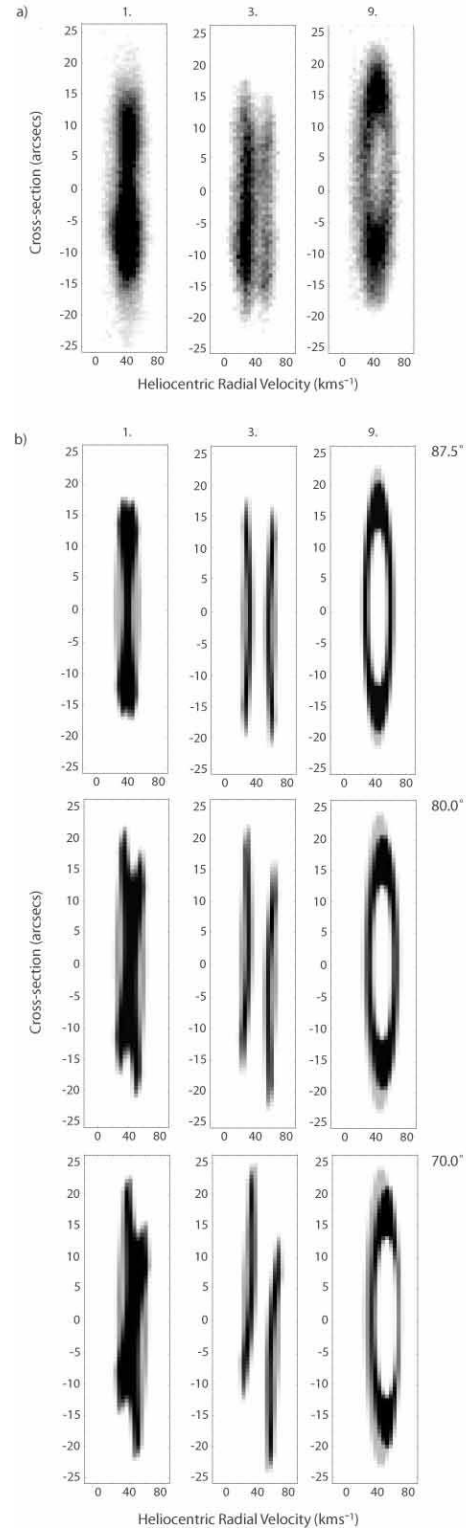}}}
\caption {(a) observed longslit spectra of Abell~63 from slits 1, 3 and 9 (from left to right). (b): convincing synthetic longslit spectra from slit positions 1, 3 and 9. Synthetic spectra are shown from models with inclinations of 87.5$^{\circ}$, 80.0$^{\circ}$ and 70.0$^{\circ}$, respectively. Expansion velocities were selected within the model that give the most convincing fit of the synthetic longslit spectra to those observed.}
\end{figure}

As a simple approximation, it is assumed that the density of the nebular rim is uniform and that the expansion velocity increases uniformly with distance from the centre of the nebula, i.e. follows a Hubble-type law. The faint nebular shell observed surrounding the bright rim of Abell~63 was modelled by a filled sphere with a uniform density. The faint shell is observed to have a surface brightness that is fainter than the rim by a factor of $\sim$2 (Fig.~1) and this has been accounted for in the model. A detailed theoretical study of the kinematics of AGB shells by \cite{2005A&A...441..573S} revealed that, in general, the outer parts of the shell expand faster than the rim, with typical velocities of 30 - 40 kms$^{-1}$ for more evolved PNe. This has been accounted for in the model, whereby the velocity of the shell increases uniformly with radius and reaches a maximum of 35 kms$^{-1}$ at the outer edge of the shell. A comparison of the deep image of Abell~63 and a two-dimensional image of the model at an inclination of 87.5$^{\circ}$ is shown in Fig.~6. 

The model shown schematically in Fig.~7. is inclined at an angle 90$^{\circ}$ - $\beta$ to the line-of-sight. The observed velocity at point X, $\rm{V}_{\rm{obsX}}$, where the expansion velocity, $\rm{V}_{\rm{expX}}$, is a maximum, is given by 

\begin{equation}
\mathrm{V}_{\mathrm{obsX}} = \mathrm{V}_{\mathrm{expX}} \cos\theta,
\end{equation}
where $\theta = \alpha - \beta$ and $\alpha$ is the angle subtending imaginary lines extending from the centre of the model to points X and Y.

The observed velocity at point Y, $\rm{V}_{\rm{obsY}}$, where the expansion velocity, $\rm{V}_{\rm{expY}}$, is at a minimum, is given by 

\begin{equation}
\mathrm{V}_{\mathrm{obsY}} = \mathrm{V}_{\mathrm{expY}} \cos\beta.
\end{equation}

The angle $\alpha$ (Fig.~7) is measured to be $\sim$47.0$^{\circ}$ for Abell~63 (Fig.~1) and thus 90$^{\circ} - \beta$ = 2.5$^{\circ}$. This gives $\theta$ $\simeq$ 44.5$^{\circ}$. 
 Fig.~4 shows that the nebular rim has an observed expansion velocity of $\sim$17 km/s. Using Equation 1, this gives a true maximum rim expansion velocity for Abell~63 of $\sim$24 kms$^{-1}$. Note that this value is governed by the assumption that the velocity increases uniformly with radius, i.e. exhibits a Hubble-type flow. 

Synthetic longslit spectra were generated by collapsing the datacube along a spatial direction orthogonal to the slit over 2 pixels and then extracting these slices. The synthetic longslit spectra were then compared with the observed \HA\ longslit spectra from Abell~63. The velocity dispersion per pixel of the model is 4.79~kms$^{-1}$ and the spatial resolution is 0.62\arcsec pixel$^{-1}$, which is consistent with the observational data in this paper. Synthetic longslit spectra were extracted from the datacube coinciding with slit positions 1, 3 and 9 (Fig.~6). 

The observed longslit spectra from slits 1, 3 and 9 are shown again in Fig.~8a and their synthetic counterparts for model inclinations 87.5$^{\circ}$, 80$^{\circ}$ and 70$^{\circ}$ are shown in Fig.~8b. The synthetic spectra extracted from the model with an inclination of 87.5$^{\circ}$ most successfully reproduces the observed velocity trends in the longslit spectra from the rim and shell of Abell~63. The close resemblance between the synthetic and observed longslit spectra supports our prediction that the nebula has the same inclination as the orbital plane of the central binary system. In addition, the results suggest that both the rim and shell exhibit a Hubble-type flow.

\section{Discussion}

A possible evolutionary model for the formation of the extended lobes in Abell~63 is considered necessary. Although the model (\S4) gives a good kinematical and morphological representation of Abell~63, it does not tell us anything about the evolution or dynamics of the nebula. We present two possibilities for the formation of the extended lobes in Abell~63: 

(1) High-speed, pressure-driven jets from the central binary system excavate a wind-driven cavity through the pre-existing AGB envelope. As the jet punctures the AGB envelope, it begins to encounter less dense material (it is assumed that the AGB envelope is less dense at the poles). The result of this would be an increase in the jet opening angle and could account for the hyperbolic shape of the rim observed in Abell~63. The jets shock-interact with remnant AGB material and provided they have sufficient momentum, the bow shock will remain collimated and produce the observed elongated lobes. The tips of the lobes would be brightest where most AGB material is swept up, forming the observed end-caps. In this model, the ionised AGB envelope that remains undisturbed by the jet would account for the faint material surrounding the bright rim (Fig.~1).

The derived kinematical ages of the lobes suggest that they are marginally older than the nebular rim by $\sim$1000~years, which is consistent with expectations that disk-generated jets should form immediately after the common envelope phase, but before the main nebular shell (i.e. the bright rim) is formed (Gon{\c c}alves et al. 2001, Soker \& Livio 1994). Alternatively, the lobes may have undergone deceleration with time as they pucture through the remnant AGB envelope. In this case, the true dynamical age of the lobes may be lower than the derived kinematical age and the lobes may be coeval with the nebular rim. 

Although there is no visible evidence that jets are still active in Abell~63, \cite{1990AJ.....99.1869S} showed that jets can only be seen during the early stages of evolution, but their cooler, denser end-caps are permanently visible.

(2) A single ballistic ejection event whereby a high-density `bullet' punctures through the AGB envelope.  After the initial eruptive event, the gas continues outwards under its own inertia in a Hubble-like flow. 

\cite{2004ASPC..313..148C} pointed out that most highly collimated PNe exhibit Hubble-like outflows with `ballistic' motions, e.g. NGC~6302 (Meaburn et al. 2005), Mz-3 (Meaburn \& Walsh 1985; Santander-Garc{\'{\i}}a et al. 2004) and NGC 6537 (Corradi \& Schwartz 1993). This is suggestive of an eruptive event that shapes the nebula very quickly rather than ongoing hydrodynamical shaping (although he states that there are exceptions to this). The lobes of Abell~63 appear very similar to the elongated lobes of Mz-3, which are shown to obey a Hubble-like velocity law; \cite{2004A&A...426..185S} concluded they are the result of a brief, eruptive formation process. 

We still face the fundamental problem of how the jets or bullets might be produced by the central binary system. The role played by close-binary central stars in the production of collimated outflows in PNe is a widely debated topic in current PNe research (e.g Garc{\'{\i}}a-Arredondo \& Frank 2004, Soker 1998). \cite{2000ApJ...538..241S} argued for the presence of a low-mass binary companion in order to account for the origin of jet-like outflows. They predicted that the very high density contrast required to form highly collimated outflows can only be produced if an accretion disk is present. The possible mechanisms for how a close-binary system might produce highly-collimated, axisymmetric PNe has been explored by \cite{1999ApJ...524..952R} and more recently by \cite{2006astro.ph..4445N}. They concluded that spin-up of the common envelope may produce a dynamo driven jet. 

The two possible mechanisms for mass transfer and resultant accretion disk formation in a binary system are wind accretion and Roche lobe overflow. Wind accretion occurs at binary separations within a few AU's and is much less restrictive to the type of companion than Roche Lobe overflow (it can be a Main Sequence star or a white dwarf). Roche lobe overflow will only occur if the primary and secondary stars are at very short separations of the order of a few solar radii and matter can be transferred onto the primary AGB star from the secondary. If sufficient angular momentum is available then the transferred matter will form an accretion disk around the primary star. This process causes a spin-up of the primary and this is thought to result in an enhanced magnetic field. The rotational motion of the accretion disk is converted into the axial motion of a jet, which may be collimated via a disk-generated magnetic field; a magnetically collimated jet has recently been directly observed by \cite{2006Natur.440...58V}. According to \cite{2006astro.ph..4445N}, Roche lobe overflow is restricted to very low-mass companions ($\le$0.3\msun). \cite{1993MNRAS.262..377P} derived the mass of the secondary in UU Sge to be 0.26\msun\ and found that the system has an orbital separation of just a few solar radii. This makes UU Sge a likely candidate for Roche lobe overflow and could account for the presence of the jet-like structures.
 
Less is known about the origin of `bullets', but it is conceivable that they are ejected in a similar mechanism to a long-lived jet, but with an enhanced mass-loss rate over a shorter time-scale. 

\section{Conclusions}

The kinematical data and morphological-kinematical modelling presented in this paper show, for the first time, that a binary central system directly affects the shaping of its nebula. The morphological-kinematical model that best reproduces the large-scale features of the longslit spectra has an inclination of 87.5$^{\circ}$; this is consistent with the well-constrained inclination of the binary orbit. The binary system is thus shown to be responsible for the formation of a jet-like structure along its polar axis.

The kinematical data confirm that the central nebula of Abell~63 has a hollow tube-like structure. The faint, collimated lobes are likely to be high-velocity outflows from the central binary system, which have excavated a cavity through the pre-existing AGB envelope. In previous work, the nebula has been described as `elliptical' (Bond \& Livio 1990), but we have shown that the morphology of Abell~63 is closer to the extremely elongated PNs M~2-9 and Mz~3. The derived kinematical ages of the lobes and rim are consistent with current models of disk-generated jets in close-binary PNe. 

The Hubble-type flow exhibited by the bright nebular rim is well-established in the morphological-kinematical modelling of the observed line profiles and imagery. The nature of the bipolar outflow in this case is unclear, it may be a continuous, low-density pressure-driven jet or a high-density momentum-driven bullet ejected from an accretion disk around the primary star. Further observations of PNe containing jet-like structures are required to gain a better understanding of the nature of jet formation in binary systems.

Further deep imaging and kinematical observations of PNe with confirmed close-binary central stars are required to investigate whether the presence of the binary influences the shaping of its nebula in a similar way in \textit{all} known cases. Only then can we make generalisations about the role played by close-binary stars in PN evolution. 

\section*{Acknowledgements}

DLM thanks PPARC for her research studentship. We would like to thank the 
staff at the San Pedro Martir observatory who helped with the observations. Thanks also to Noam Soker for useful discussions.

\label{lastpage}


\begin{thebibliography}{}

\bibitem[Balick et al.(1987)]{1987AJ.....94.1641B} Balick, B., Preston, 
H.~L., Icke, V.,\ 1987, \aj, 94, 1641 

\bibitem[Bell et al.(1994)]{1994MNRAS.270..449B} Bell, S.~A., Pollacco, 
D.~L., Hilditch, R.~W.,\ 1994, \mnras, 270, 449
 
\bibitem[Bond et al.(1978)]{1978ApJ...223..252B} Bond, H.~E., Liller, W., 
 Mannery, E.~J.,\ 1978, \apj, 223, 252

\bibitem[Bond \& Livio(1990)]{1990ApJ...355..568B} Bond, H.~E., Livio, 
M.,\ 1990, \apj, 355, 568 

\bibitem[Corradi \& Schwarz(1993)]{1993A&A...269..462C} Corradi, R.~L.~M.,
 Schwarz, H.~E.,\ 1993, \aap, 269, 462 

\bibitem[Corradi(2004)]{2004ASPC..313..148C} Corradi, R.~L.~M.,\ 2004, ASP 
Conf.~Ser.~313: Asymmetrical Planetary Nebulae III: Winds, Structure and 
the Thunderbird, 313, 148

\bibitem[Frank(2005)]{2005AIPC..804...81F} Frank, A.,\ 2005, AIP 
Conf.~Proc.~804: Planetary Nebulae as Astronomical Tools, 804, 81

\bibitem[Garc{\'{\i}}a-Arredondo \& Frank(2004)]{2004ApJ...600..992G} 
Garc{\'{\i}}a-Arredondo, F., Frank, A.,\ 2004, \apj, 600, 992 

\bibitem[Garc{\'{\i}}a-Segura et al.(1999)]{1999ApJ...517..767G} 
Garc{\'{\i}}a-Segura, G., Langer, N., R{\'o}{\.z}yczka, M., \& Franco, J.\ 
1999, \apj, 517, 767

 \bibitem[Gill \& O'Brien(1999)]{1999MNRAS.307..677G} Gill, C.~D., O'Brien, T.~J.,\ 1999, \mnras, 307, 677

\bibitem[Gon{\c c}alves et al.(2001)]{2001ApJ...547..302G} Gon{\c c}alves, 
D.~R., Corradi, R.~L.~M., \& Mampaso, A.\ 2001, \apj, 547, 302

\bibitem[Kahn \& West(1985)]{1985MNRAS.212..837K} Kahn, F.~D., West, 
K.~A.,\ 1985, \mnras, 212, 837 

\bibitem[Livio et al.(1979)]{1979MNRAS.188....1L} Livio, M., Salzman, J., 
 Shaviv, G.,\ 1979, \mnras, 188, 1 

\bibitem[Lloyd et al.(1993)]{1993MNRAS.265..457L} Lloyd, H.~M., Bode, 
M.~F., O'Brien, T.~J., Kahn, F.~D.,\ 1993, \mnras, 265, 457

\bibitem[De Marco et al. (2004)]{2004ApJ...602L..93D} De Marco, O., Bond, H.E., Harmer, D., Fleming, A.J., \ 2004, \apj, 602, L93

\bibitem[Mastrodemos \& Morris(1998)]{1998ApJ...497..303M} Mastrodemos, N., 
 Morris, M.,\ 1998, \apj, 497, 303 

\bibitem[Meaburn \& Walsh(1985)]{1985MNRAS.215..761M} Meaburn, J., 
Walsh, J.~R.,\ 1985, \mnras, 215, 761

\bibitem[Meaburn et al.(2003)]{2003RMxAA..39..185M} Meaburn, J., L{\'o}pez, 
J.~A., Guti{\'e}rrez, L., Quir{\'o}z, F., Murillo, J.~M., Vald{\'e}z, J., 
 Pedrayez, M.,\ 2003, RMxAA, 39, 
185

\bibitem[Meaburn et al.(2005)]{2005AJ....130.2303M} Meaburn, J., L{\'o}pez, 
J.~A., Steffen, W., Graham, M.~F., Holloway, A.~J.,\ 2005, \aj, 130, 2303 

\bibitem[Morris(1987)]{1987PASP...99.1115M} Morris, M.,\ 1987, \pasp, 99, 
1115  

\bibitem[Nordhaus \& Blackman(2006)]{2006astro.ph..4445N} Nordhaus, J., 
Blackman, E.~G.,\ 2006, preprint (astro-ph/0604445)

\bibitem[Pollacco \& Bell(1993)]{1993MNRAS.262..377P} Pollacco, D.~L., 
Bell, S.~A.,\ 1993, \mnras, 262, 377

\bibitem[Pollacco \& Bell(1994)]{1994MNRAS.267..452P} Pollacco, D.~L., 
Bell, S.~A.,\ 1994, \mnras, 267, 452   

\bibitem[Pollacco \& Bell(1997)]{1997MNRAS.284...32P} Pollacco, D.~L., 
Bell, S.~A.,\ 1997, \mnras, 284, 32 

\bibitem[Reyes-Ruiz \& L{\'o}pez(1999)]{1999ApJ...524..952R} Reyes-Ruiz, 
M., \& L{\'o}pez, J.~A.\ 1999, \apj, 524, 952 

\bibitem[Sahai \& Trauger(1998)]{1998AJ....116.1357S} Sahai, R., 
Trauger, J.~T.,\ 1998, \aj, 116, 1357 

\bibitem[Santander-Garc{\'{\i}}a et al.(2004)]{2004A&A...426..185S} 
Santander-Garc{\'{\i}}a, M., Corradi, R.~L.~M., Balick, B., Mampaso, A.,\ 
2004, \aap, 426, 185 
 
\bibitem[Schoenberner(1981)]{1981A&A...103..119S} Schoenberner, D.,\ 1981, 
\aap, 103, 119 

\bibitem[Schoenberner(1983)]{1983ApJ...272..708S} Schoenberner, D.,\ 1983, 
\apj, 272, 708 

\bibitem[Sch{\"o}nberner et al.(2005)]{2005A&A...441..573S} 
Sch{\"o}nberner, D., Jacob, R., Steffen, M.,\ 2005, \aap, 441, 573

\bibitem[Soker(1990)]{1990AJ.....99.1869S} Soker, N.,\ 1990, \aj, 99, 1869

\bibitem[Soker \& Livio(1994)]{1994ApJ...421..219S} Soker, N., Livio, 
M.,\ 1994, \apj, 421, 219

\bibitem[Soker(1998)]{1998ApJ...496..833S} Soker, N.,\ 1998, \apj, 496, 833

\bibitem[Soker \& Rappaport(2000)]{2000ApJ...538..241S} Soker, N.,
Rappaport, S.,\ 2000, \apj, 538, 241

\bibitem[Soker(2002)]{2002ApJ...568..726S} Soker, N.,\ 2002, \apj, 568, 726 

\bibitem[Soker \& Lasota (2004)]{2004AA...422.1039S} Soker, N., Lasota, J. ~P.,\ 2004, A\&A, 422, 1039

\bibitem[Soker(2006)]{2006PASP...118..260S} Soker, N.,\ 2006, \pasp, 118, 260

\bibitem[Sorensen \& Pollacco (2004)]{2004ASPC..313..515S} Sorensen, P., Pollacco, D.L., \ 2004, In
{\it Asymmetrical Planetary Nebulae III: Winds, Structure and the Thunderbird}, Edited by M. Meixner, J.H. Kastner, B.Balick and N.Soker. ASP Conf. Proc., Vol. 313. San Francisco, p.515.

\bibitem[Steffen \& L{\'o}pez(1998)]{1998ApJ...508..696S} Steffen, W.,
L{\'o}pez, J.~A.,\ 1998, \apj, 508, 696 

\bibitem[Vlemmings et al.(2006)]{2006Natur.440...58V} Vlemmings, W.~H.~T., 
Diamond, P.~J., Imai, H.,\ 2006, Nat, 440, 58 
  

\end{thebibliography}
\end{document}